\documentclass[aps,prc,reprint,longbibliography,superscriptaddress,floatfix,nofootinbib]{revtex4-2}

\usepackage{graphicx}
\graphicspath{{./}{figures/}}
\usepackage{booktabs}
\usepackage{amsmath}
\usepackage{amssymb}
\usepackage{bm}
\usepackage{xcolor}
\usepackage{hyperref}
\hypersetup{
  colorlinks=true,
  linkcolor=black,
  citecolor=black,
  urlcolor=black,
  filecolor=black,
  pdfborder={0 0 0},
  breaklinks=true
}
\begin{document}

\title{Spin-Dependent Nucleon--Nucleus Interactions Constrained by Neutron Observables and Their Impact on Near-Barrier Proton Fusion}

\author{Kyoungsu Heo}
\thanks{Corresponding author.}
\affiliation{Department of Physics and OMEG Institute, Soongsil University, Seoul 06978, Korea}
\author{Myung-Ki Cheoun}
\affiliation{Department of Physics and OMEG Institute, Soongsil University, Seoul 06978, Korea}
\author{K.\ Hagino}
\affiliation{
Department of Physics, Kyoto University, Kyoto 606-8502, Japan} 
\affiliation{Institute for Liberal Arts and Sciences, Kyoto University, Kyoto 606-8501, Japan}
\affiliation{
RIKEN Nishina Center for Accelerator-based Science, RIKEN, Wako 351-0198, Japan
}

\date{October 26, 2026}

\begin{abstract}
We investigate the role of spin-dependent nucleon--nucleus interactions in nuclear reactions. 
To this end, we use neutron spin observables to constrain the dominant central spin-spin form factors and then apply the corresponding like-channel interactions to near-barrier fusion in the $p+{}^{93}$Nb system. The interactions are constructed within a folding framework based on a finite-range effective nucleon-nucleon force and organized in terms of
radial form factors associated with their spin-spin, tensor, and spin-orbit components. Neutron spin observables in the $n$ + $^{27}$Al, $^{59}$Co, and $^{93}$Nb
target systems are analyzed within a distorted-wave Born approximation (DWBA) framework {to examine the sign and scale of the central spin-spin contribution and the assembled operator conventions.} The calculation reproduces the observed sign systematics of the neutron spin observables for the three targets, {{consistent with the dominant central spin-spin contribution in the adopted model.}} 
The unlike-channel (neutron--proton) interaction constrained by neutron scattering is then reconstructed for the corresponding like-channel (proton--proton) interaction and applied to a coupled-channels description of near-barrier fusion for the $p+{}^{93}$Nb system. The {retained real} spin-dependent interactions lead only to a weak modification of the effective barrier and {a very small change in the unpolarized} fusion cross section {within the present model.} The present work thus connects the neutron-scattering constraints on the operator conventions with the fusion calculation in the proton channel, and {estimates the scale} of the corresponding real spin-dependent correction in near-barrier fusion.
\end{abstract}

\maketitle

\section{Introduction}

While a folded potential for low-energy reaction calculations usually has a purely 
central form, 
spin-dependent components of the nucleon--nucleus interaction 
induce a noncentral character. These terms, arising from the spin-dependent nucleon--nucleon (NN) interaction, can in principle influence reaction observables in both scattering and fusion processes \cite{McAbee1990,McAbee1986,Clarke1974,Sakuragi1998,Satchler1971b,Ishikawa2005,Cunningham2013,Petrovich1978}. While their role has been explored in polarized-neutron scattering, their quantitative impact on proton-induced reactions near the Coulomb barrier remains less well understood.

A central difficulty in assessing spin-dependent effects in proton reactions is that near-barrier fusion observables are dominated by competing mechanisms such as barrier penetration, channel couplings, and absorption. As a result, isolating the specific contribution of spin-dependent interactions is nontrivial. In contrast, polarized-neutron observables provide a more direct probe of the spin-dependent interaction, because the spin-dependent observables are constructed directly from the corresponding spin-dependent form factors~\cite{McAbee1990,McAbee1986}. The corresponding data sets for the $^{27}$Al, $^{59}$Co, and $^{93}$Nb targets~\cite{Stamp1967,Nagamine1970,Heeringa1977,Gould1986,Soderstrum1992,Heeringa1989}, together with modern reaction-model studies~\cite{Nagadi2004}, therefore provide a useful basis for constraining the sign and {scale} of the dominant central spin-spin component and checking the assembled operator conventions.

The aim of this paper is to clarify the role of spin-dependent interactions in proton reactions. To this end, 
 we adopt a two-step strategy. First, we 
use neutron spin observables
 within a distorted-wave Born approximation (DWBA) description to
 constrain the dominant central spin-spin parts of the folded radial form factors and examine the operator structures and conventions. Second, using the DWBA analysis of neutron scattering, we reconstruct the corresponding interaction in the like-nucleon (proton-proton) channel and use it to evaluate its impact on near-barrier proton fusion in $p+{}^{93}$Nb within a coupled-channels framework~\cite{BalantekinTakigawa1998,Dasgupta1998,HaginoTakigawa2012}. The folded interaction is generated with the finite-range Tokyo-4-Yukawa (T4Y) effective interaction~\cite{Hosaka1985,McAbee1986,McAbee1990}.
For the neutron DWBA validation, the distorted waves are generated with
the CH89 global optical model potential, including its central and
conventional projectile spin-orbit terms. For the application to fusion in the proton channel, the reference barrier is constructed from the real
central component of the global proton optical potential with the geometry parameters of Table~VI of Ref.~\cite{Varner1991}, together with the Coulomb and centrifugal terms. This potential is used as a fixed reference baseline rather than as an optimized fusion potential.
This comparison provides an estimate of the magnitude of the corresponding like-channel
spin-dependent correction in near-barrier proton fusion. The central question addressed here is therefore how large the neutron-constrained real spin-dependent correction remains when it is propagated into a near-barrier proton-fusion calculation.

The paper is organized as follows. Section~\ref{sec:framework} presents
the theoretical framework, operator conventions, and observable construction.
Section~\ref{sec:neutron} discusses the constraints from neutron spin
observables. Section~\ref{sec:proton} examines the impact of the
reconstructed interaction on near-barrier proton fusion in $p+{}^{93}$Nb.
Section~\ref{sec:discussion} discusses the remaining discrepancy and the role
of absorption effects. Conclusions are given in Sec.~\ref{sec:conclusions}.

\section{Theoretical Framework}
\label{sec:framework}

\subsection{Spin-dependent interaction model}

The analysis of neutron scattering follows the operator basis used in the
previous folding-model studies of polarized-neutron scattering~\cite{McAbee1990,McAbee1986}, which explicitly employed
the central spin-spin, tensor, and spin-orbit sectors. The folded radial functions are organized into the form factors $F_{10}(R)$, $F_{12}(R)$, and $F_{32}(R)$, where $F_{ik}$ denotes the radial form factor associated with the target-spin-dependent operator of spin rank $i$ and spatial rank $k$ (see Eq. (\ref{eq:interaction_decomposition}) below). These form factors enter the cross-section {components} $\sigma_{10}$, $\sigma_{12}$, and $\sigma_{32}$ in the distorted-wave Born approximation (DWBA). 

The neutron-scattering analysis and the proton-fusion application share the same $S_{ik}$ operator basis and the same sign, normalization, and conventions used to construct the operators, while the folded radial form factors are evaluated separately in the unlike-nucleon (i.e., neutron-proton) and like-nucleon (i.e., proton-proton and neutron-neutron)
channels. For the neutron scattering analysis, the folded spin-dependent form factors enter
through DWBA transition matrix elements with distorted waves generated by the
optical potential. For the proton application, the corresponding like-channel form factors are used in a fixed-$(J,S,l)$ coupled-channels calculation, where $J$ is the total angular momentum of the system while $S$ and $l$ are
the total spin and the relative orbital angular momentum between the colliding nuclei, respectively.

The nucleon-nucleus interaction is decomposed schematically as
\begin{equation}
\begin{aligned}
U(\bm R)
&=
U_{\mathrm{opt}}(\bm R)
\\
&\quad-
\sum_{(i,k)=(1,0),(1,2),(3,2)}
F_{ik}(R)\,S_{ik}(\hat{\bm R},\bm s_p,\bm I),
\end{aligned}
\label{eq:interaction_decomposition}
\end{equation}
where the minus sign in Eq.~(\ref{eq:interaction_decomposition}) follows the convention of Refs.~\cite{McAbee1990,McAbee1986} and is used consistently in both the neutron DWBA reduction and the proton coupled-channels calculation. $\bm R$ is the projectile-target separation vector, $\hat{\bm R}$ is its direction, $\bm s_p$ is the projectile-spin operator, and $S_{ik}$ denotes the tensor product of the spin coupled by target spin rank $i$ and projectile spin and the spatial rank $k$ as shown below
\begin{equation}
S_{ik} \equiv N_{ik} {[{\bm I}^{(i)} \times {\bm s_p}]}^{(k)} \cdot [\hat{\bm R}]^{(k)}~,
\end{equation}
where $N_{ik}$ is the normalization factor, ${\bm I}^{(i)}$ is the target spin tensor, the index $i$ labels the
spin rank in target-spin space, and the index $k$ labels the external
spatial rank associated with the projectile--target separation direction.
The tensor $[\hat{\bm R}]^{(k)}$ denotes the rank-$k$ irreducible
spatial tensor constructed from $\hat{\bm R}$. {We use $[\hat{\bm R}]^{(k)}_q=\sqrt{4\pi/(2k+1)}Y_{kq}(\hat{\bm R})$ and fix $N_{ik}$ by $\langle I,I;s_p,s_p|S_{ik}(\hat{\bm z})|I,I;s_p,s_p\rangle=1$. For nucleon spins, $\bm s=\bm\sigma/2$ in units of $\hbar$, where $\bm\sigma$ denotes the Pauli matrices.}

The internal coordinate of the valence nucleon in the target nucleus
enters only through the folded radial form factors. In the present work, only the spatial ranks $k=0$ and $k=2$ are retained. 

In the present operator basis, the $(1,0)$ part is a spatially scalar spin-dependent component corresponding to $F_{10}$, and the $(1,2)$ and $(3,2)$ parts are spatially rank-2, quadrupole-like spin-dependent form factors, $F_{12}$ and $F_{32}$.

The quantity $U_{\mathrm{opt}}$ in Eq. (\ref{eq:interaction_decomposition})
is the optical potential independent of target spin, while $F_{ik}$ carries the target-spin-dependent interaction.
In the neutron DWBA calculation, $U_{\mathrm{opt}}$ includes the central and
conventional projectile spin-orbit terms of the optical potential introduced
above. The latter term belongs to the optical
potential that generates the distorted waves, whereas the direct spin-orbit
parts $F_{ik}^{(ls)}$ are folded target-spin-dependent contributions to
$F_{10}$, $F_{12}$, and $F_{32}$.
This separation prevents double counting.
 For the folded interaction, we adopt the T4Y finite-range spin-dependent effective nucleon-nucleon interaction~\cite{Hosaka1985,McAbee1986,McAbee1990,SatchlerLove1979}. M3Y-type alternatives discussed in the same literature are useful comparison points: their separate direct and exchange parts can differ substantially, whereas the direct-plus-exchange totals are often much closer once the exchange term is included~\cite{McAbee1990,Bertsch1977,Anantaraman1983,Love1978}. For that reason we keep the T4Y interaction inputs in the present work. The neutron scattering analysis uses the unlike-nucleon channel, while the proton application uses the corresponding like-nucleon channel.

For the neutron scattering, the radial form factor is decomposed as
\begin{equation}
F_{ik}(R)=F_{ik}^{(ss)}(R)+F_{ik}^{(t)}(R)+F_{ik}^{(ls)}(R),
\label{eq:Fik_split}
\end{equation}

The assembled spin-dependent
nucleon--nucleus term is denoted by $U_{ik}(R)=-F_{ik}(R)S_{ik}$.
In this notation, $F_{ik}^{(ss)}$ contains the direct-plus-exchange central
spin-spin contribution, with the exchange part evaluated through the local
single-nucleon-knockout prescription used in
Refs.~\cite{McAbee1990,McAbee1986}. The tensor and spin-orbit contributions,
$F_{ik}^{(t)}$ and $F_{ik}^{(ls)}$, are evaluated using only the direct
components of the folding integrals; their exchange parts would require
nonlocal or momentum-dependent reductions beyond the local form-factor
treatment adopted here~\cite{Cunningham2013}. They are therefore omitted 
not because they vanish, but because their consistent treatment lies outside the present local form-factor scheme. This separation allows the
central spin-spin, tensor, and spin-orbit sectors to be examined separately.

In this paper, following Refs. \cite{McAbee1986,McAbee1990}, we assume that the 
target nuclei consist of a spin-zero core nucleus and a valence nucleon.
Using the standard angular-momentum reduction, the radial form factors are factorized as
\begin{align}
F_{ik}^{(ss)}(R)
&=
\frac{3}{2\sqrt{2}}\,
\mathcal A_{ik}(I,l_t) U_k^{(ss)} (R),
\label{eq:Fik_ss_general}
\\
F_{ik}^{(t)}(R)
&= - 4
\sum_n C_{ikn}^{(t)}(I,l_t)\, U_{kn}^{(t)}(R),
\label{eq:Fik_t_general}
\\
F_{ik}^{(ls)}(R)
&=
\sum_{m\in\{0,2,3\}} C_{ikm}^{(ls)}(I,l_t)\,W_m^{(ls)}(R),
\label{eq:Fik_ls_general}
\end{align}
%

Here $U_k^{(ss)}(R)$ denotes the direct-plus-exchange folded radial component
of the central spin-spin interaction. Its direct part is obtained by folding
$v_k^{(ss)}(R,\alpha r_t)$, the multipole component of the spin-spin
interaction between the projectile nucleon and the valence nucleon in the
target, with $u_{l_t}^2(r_t)$, where $u_{l_t}(r_t)$ is the normalized
bound-state radial wave function of the same valence nucleon. The local
exchange part is also included. The detailed prescription is summarized in
Appendix~\ref{app:fik}.

The orbital angular momentum of the valence nucleon is denoted by
$l_t$, $r_t$ is measured from the target center, and
$\alpha=A_t/(A_t+1)$ is the recoil factor. 

The density entering the folding integrals (see Eq.~(\ref{eq:uss_direct})) is the single-particle density of
the valence nucleon in the target nucleus, a proton for the three targets
considered here, rather than the total matter or charge density. It is
constructed from the normalized bound-state radial wave function as
\begin{equation}
\rho_v(r_t)=\frac{1}{4\pi}\left[\frac{u_{l_t}(r_t)}{r_t}\right]^2,
\end{equation}
with
\begin{equation}
\int_0^\infty u_{l_t}^2(r_t)\,dr_t=1 .
\end{equation}
The adopted valence-proton orbits are $1d_{5/2}$, $1f_{7/2}$, and $1g_{9/2}$,
corresponding to $l_t=2$, 3, and 4, for $^{27}$Al, $^{59}$Co, and
$^{93}$Nb, respectively.
The bound-state wave functions are obtained from a Woods--Saxon potential with
$R_0=r_0A_c^{1/3}$, $r_0=1.20$~fm, and $a_0=0.65$~fm, including the
spin-orbit and Coulomb terms, where $A_c$ is the mass number of the core nuclei. The central depth is
adjusted to reproduce the empirical proton separation energy. With the adopted
values of $S_p=8.271$, $7.364$, and $6.043$~MeV for
$^{27}$Al, $^{59}$Co, and $^{93}$Nb, respectively~\cite{Wang2021AME2020}, 
the depth parameter is found to be 
$V_0=-56.588$, $-56.657$, and
$-60.101$~MeV for
$^{27}$Al, $^{59}$Co, and $^{93}$Nb, respectively.

Detailed forms of the folded tensor and spin-orbit terms, $U_{kn}^{(t)}(R)$ and $W_m^{(ls)}(R)$, as well as the relevant angular momentum couplings, $C_{ikn}^{(t)}(I,l_t)$ and $C_{ikm}^{(ls)}(I,l_t)$, are provided in Eqs.~(\ref{eq:Ukn_definition}), (\ref{eq:Cikn_tensor}), and (\ref{eq:W0ls})--(\ref{eq:Fik_ls_appendix}) in Appendix~\ref{app:fik}. The reductions of the folding potential and observables used here follow the standard approach for polarized neutrons and the conventional angular-momentum algebra summarized in Refs.~\cite{McAbee1990,McAbee1986,BrinkSatchler1993,Edmonds1960,Krane1973,HorieSasaki1961,BrievaRook1977a,BrievaRook1977b,SatchlerLove1979,Varner1991}. We keep only the terms needed for the present neutron scattering analysis and proton fusion application. For example, the angular momentum coupling for the spin-spin interaction in Eq.~(\ref{eq:Fik_ss_general}) is given as
\begin{widetext}
\begin{equation}
\mathcal A_{ik}(I,l_t)
\equiv
{\frac{(-1)^{l_t+1}(2l_t+1)}{\sqrt{3}}}
(\hat i\,\hat I\,\hat k)^2
\begin{Bmatrix}
I & \tfrac12 & l_t\\
I & \tfrac12 & l_t\\
i & 1 & k
\end{Bmatrix}
\begin{pmatrix}
l_t & k & l_t\\
0 & 0 & 0
\end{pmatrix}
\begin{pmatrix}
i & 1 & k\\
0 & 0 & 0
\end{pmatrix}
\begin{pmatrix}
I & I & i\\
I & -I & 0
\end{pmatrix},
\label{eq:Aik}
\end{equation}
\end{widetext}
where the parentheses and braces denote the standard Wigner $3j$ and $9j$ symbols of the angular-momentum algebra. 
Eqs.~(\ref{eq:Fik_split})-(\ref{eq:Fik_ls_general}) connect the operator basis to the radial form factors used in the calculated observables.

For the central spin-spin contribution alone, the common radial factor
$U_2^{(ss)}(R)$, the valence orbital $l_t$, and the spatial geometrical
factor are identical for $F_{12}^{(ss)}$ and $F_{32}^{(ss)}$. The ratio is
therefore fixed only by the $i$-dependent Wigner recoupling factors. 
For $^{93}$Nb, which has the stretched $g_{9/2}$ valence configuration, this gives
$F_{32}^{(ss)}=(3/2)F_{12}^{(ss)}$. Moreover, in this case, the closed-form coefficients are simplified substantially~\cite{McAbee1986}.

Writing the direct-plus-exchange central spin-spin 
folding potentials of spatial ranks $k=0$ and $k=2$ as $U^{(ss)}_0$ and $U^{(ss)}_2$, the tensor single-folding terms as $U_{20}^{(t)}$, $U_{22}^{(t)}$, and $U_{24}^{(t)}$, and the spin-orbit pieces as $U_{10}^{(ls)}$, $F_{12}^{(ls)}$, and $F_{32}^{(ls)}$, the radial form factors can be written in the following compact form~\cite{McAbee1986}; 

\begin{widetext}
\begin{align}
F_{10}(R) &= a\,U^{(ss)}_0(R) + b\,U^{(ls)}_{10}(R) + c\,{T_{02}}(R),
\label{eq:fik_reduced_f10} \\
F_{12}(R) &= c\,U^{(ss)}_2(R) - c\,T_{22}(R) + 2a\,T_{20}(R) + F_{12}^{(ls)}(R),
\label{eq:fik_reduced_f12} \\
F_{32}(R) &= \frac{3}{2}c\,U^{(ss)}_2(R) - \frac{2}{7}\left(\frac{3}{2}c\right)T_{22}(R) - \frac{9b_{32}}{7a_{32}}\left(\frac{3}{2}c\right)T_{24}(R) + F_{32}^{(ls)}(R),
\label{eq:fik_reduced_f32}
\end{align}
\end{widetext}
 with $a$=1/4, $b$=1, $c=-2/11$, $a_{32}=-13$, and $b_{32}=-6$. For the tensor sector, $T_{kn}(R) \equiv -4U_{kn}^{(t)}(R)$ denotes 
the rescaled folded tensor radial component used in the assembled form factors. 
Appendix~\ref{app:fik} provides the explicit $(k,n)$ tensor notation used in the assembled form factors.
In the compact notation of Eqs.~(\ref{eq:fik_reduced_f10})--(\ref{eq:fik_reduced_f32}), the tensor labels
follow the original shorthand notation. In the explicit notation used in
Appendix~\ref{app:fik}, the tensor contribution in $F_{10}$ corresponds to
$T_{02}$, the $2a$ tensor contribution in $F_{12}$ corresponds to $T_{20}$,
and the remaining rank-2 tensor components are $T_{22}$ and $T_{24}$.
For the spin-orbit part, the compact notation is related to the
explicit notation in Appendix~\ref{app:fik} by $U_{10}^{(ls)}=-W_0^{(ls)}$,
$F_{12}^{(ls)}=bW_2^{(ls)}$, and
$F_{32}^{(ls)}=c_{32}^{(ls)}W_3^{(ls)}$.

To state the normalization and sign convention {for the parallel geometry}, we quote the standard
no-spin-orbit-distortion reduction used in Refs.~\cite{McAbee1990,McAbee1986}:
\begin{widetext}
\begin{equation}
\sigma_{ik}^{ss}(E)
=
\frac{4\pi}{\sqrt{2}\,k_p E}
\left[
\hat i\,\hat I
\begin{pmatrix}
I & I & i\\
I & -I & 0
\end{pmatrix}
\right]^{-1}
\sum_{l,l'}
(-1)^{(l-l')/2}
(\hat l\,\hat l')^2
\begin{pmatrix}
l & l' & k\\
0 & 0 & 0
\end{pmatrix}^{{2}}
\Im\!\int_0^\infty dr\,
u_{l'}(k_p,r)\,
F_{ik}^{(ss)}(r)\,
u_l(k_p,r).
\label{eq:sigmaik_mcabee}
\end{equation}
\end{widetext}
This formula shows how the radial form factors are folded once more
with the distorted radial waves $u_l(k_p,r)$. Here $E$ is the incident center-of-mass energy, $k_p$ is the projectile wave number, and $\Im$ denotes the imaginary part of the radial integral. {Notice that the figures for neutron scattering are plotted as functions of $E_{\rm lab}$. The prefactor is evaluated with $E=E_{\rm cm}=A_tE_{\rm lab}/(A_t+1)$.}
In the no-spin-orbit-distortion limit, the relative size of the two
$i=k\pm1$ components reads
\begin{equation}
\frac{\sigma_{k+1,k}^{ss}}{\sigma_{k-1,k}^{ss}}
=
\frac{k+1}{k}
\left[
\frac{(2k-1)(2I+k+2)(2I+k+1)}{(2k+3)(2I-k+1)(2I-k)}
\right]^{1/2},
\label{eq:sigmaik_ratio}
\end{equation}
which is useful for a consistency check on the relative size of the two $i=k\pm1$ components.
Equation~(\ref{eq:sigmaik_mcabee}) is used as a reference reduction formula for fixing the normalization and sign convention of $\sigma_{ik}^{ss}$; this analytic reduction corresponds to the limit in which the projectile spin-orbit distortion is omitted. In the numerical calculation,
the distorted waves are generated with the optical potential including its central and projectile spin-orbit terms, and the DWBA radial matrix elements are
evaluated with the corresponding $j=l\pm1/2$ distorted partial waves.

In brief, the folding construction of the spin-dependent form factors
combines two elements. First, the valence-nucleon density is constructed from the bound-state single-particle wave function obtained by solving the radial Schr\"odinger equation with a Woods--Saxon potential, including the spin-orbit term and the Coulomb term for the proton case. Second, the direct and exchange kernels of the central spin-spin term are generated within the same T4Y folding-model framework, which is an energy-independent finite-range interaction. In particular, the exchange term is treated explicitly through the published local single-nucleon-knockout exchange approximation, rather than being absorbed into an effective direct term. This means that the practical energy dependence of the folded central spin-spin potential enters mainly through the local-exchange part.

We note that the neutron scattering analysis and proton fusion application, therefore, share the same operator decomposition, but they use separately reconstructed unlike- and like-channel form factors. {The neutron analysis computes $F_{ik}$ and evaluates the corresponding components $\sigma_{ik}$ with one-channel distorted waves in DWBA.} The proton application uses the corresponding like-channel form factors in a truncated fixed-$(J,S,l)$ coupled-channels calculation and evaluates the fusion cross section from the resulting penetrabilities.

\subsection{Observable construction and geometry}
\label{sec:observable_geometry}

{With projectile spin-orbit distortion, the spin-dependent cross section is obtained from the full forward DWBA amplitude for each polarization geometry as}
\begin{equation}
\begin{aligned}
&{\frac{\sigma_{ss}}{P_tP_b}
=\frac{t_{10}}{P_t}\left[\sigma_{10}+\sigma_{12}\right]
+\frac{t_{30}}{P_t}\sigma_{32}.}
\end{aligned}
\label{eq:sigmass_reduction}
\end{equation}
{Here $P_t$ and $P_b$ are the target and beam polarizations, and $t_{10}$ and $t_{30}$ are the target statistical tensors. Each component $\sigma_{ik}$ for the specified geometry is {normalized to a unit value of the corresponding target statistical tensor}.}

{In the limit without projectile spin-orbit distortion, the angular dependence reduces to the Legendre polynomial $P_k(\cos\theta_I)$, where $\theta_I$ is the angle between the polarization axis and the beam direction. In this limit $P_0=1$, $P_2(1)=1$, and $P_2(0)=-1/2$~\cite{McAbee1986,McAbee1990}.}
In the following, the perpendicular and parallel geometries correspond to
the transverse and longitudinal spin orientations, respectively, in the
experimental notation.
{The following expressions use the parallel components $\sigma_{ik}$:}
\begin{eqnarray}
\left.\frac{\sigma_{ss}}{P_t P_b}\right|_{\parallel}
&=&
\frac{t_{10}}{P_t}\,\sigma_{10}
+\frac{t_{10}}{P_t}\,\sigma_{12}
+\frac{t_{30}}{P_t}\,\sigma_{32}, \\
\left.\frac{\sigma_{ss}}{P_t P_b}\right|_{\perp}
&=&
\frac{t_{10}}{P_t}\,\sigma_{10}
-\frac12
\left[
\frac{t_{10}}{P_t}\,\sigma_{12}
+\frac{t_{30}}{P_t}\,\sigma_{32}
\right].
\label{eq:sigmass_perpendicular_appendix}
\end{eqnarray}
{The experimental comparisons use Eq.~(\ref{eq:sigmass_reduction}) with projectile spin-orbit distortion; Eq.~(\ref{eq:sigmass_perpendicular_appendix}) serves only as a check with all its components evaluated without that distortion~\cite{McAbee1990}.}

For each target, the magnetic-substate probabilities are taken in the
thermal-orientation form
$p_M(x)=\exp(xM)/\sum_{M=-I}^{I}\exp(xM)$, where the orientation parameter
$x$ is fixed by the input target polarization
$P_t=\sum_M M p_M/I$. The statistical tensors are evaluated as
$t_{i0}=\hat I\hat i\sum_{M=-I}^{I}(-1)^{I-M}
\begin{pmatrix} I&I&i\\ M&-M&0 \end{pmatrix}p_M$.
The $P_t$ values in Table~\ref{tab:orientation_factors} are the
spin-temperature inputs used to compute the statistical-tensor ratios
{for $B=7$~T and $T=10$~mK~\cite{McAbee1990}}.
They should not be identified with the effective target polarizations
quoted for individual experimental runs.
The comparison is made directly to the normalized observable
$\sigma_{ss}/(P_tP_b)$ reported in the neutron scattering
literature~\cite{Gould1986,Heeringa1977,Heeringa1989,Soderstrum1992}.
Thus $P_b$ is not introduced as an independent theory input.

\begin{table}[t]
\caption{\label{tab:orientation_factors}Target-orientation factors used in
the neutron scattering observables. {The listed $P_t$ values specify spin-temperature inputs at $B=7$~T and $T=10$~mK~\cite{McAbee1990}.} The same $P_t$,
$t_{10}/P_t$, and $t_{30}/P_t$ values are used for the parallel and
perpendicular geometries.}
\begin{ruledtabular}
\begin{tabular}{ccccc}
Nucleus & $I$ & $P_t$ & $t_{10}/P_t$ & $t_{30}/P_t$ \\
\hline
$^{27}$Al & $5/2$ & $0.428$ & $1.46$ & $0.0709$ \\
$^{59}$Co & $7/2$ & $0.500$ & $1.53$ & $0.127$ \\
$^{93}$Nb & $9/2$ & $0.640$ & $1.57$ & $0.270$ \\
\end{tabular}
\end{ruledtabular}
\end{table}

\section{Constraints from Neutron Spin Observables}
\label{sec:neutron}

In this section, we use the spin observables in neutron scattering to
{examine the sign and scale of the central spin-spin contribution} and to
check the conventions used to construct the operators. The emphasis is not on a parameter fit, but on whether the constructed interaction reproduces the characteristic sign and structure of the measured observables. Here the sign means 
 that of the assembled neutron spin-spin observable, after the radial form factors, statistical tensors, and geometry factors are combined, rather than the sign of a single radial term.

Before presenting the neutron scattering results as a benchmark for proton fusion, we summarize four conventions used throughout the analysis. First, the DWBA observable reduction uses the normalization factor $\hat I=\sqrt{2I+1}$. Second, in the unlike-channel case the direct term carries the minus sign required by the angular-momentum recoupling algebra. Third, the $F_{12}$ form factor includes the tensor and spin-orbit terms following Ref.~\cite{McAbee1990}. Fourth, the $F_{32}$ form factor includes the $(2,4)$ tensor component through $T_{24}$ with the same fixed angular-momentum normalization used for the other tensor terms. This inclusion moderately affects the radial form factors and the neutron spin observables. 

\subsection{Form factors for \texorpdfstring{$^{93}\mathrm{Nb}$}{93Nb}}

\begin{figure}
  \centering
  \includegraphics[width=0.48\textwidth,keepaspectratio]{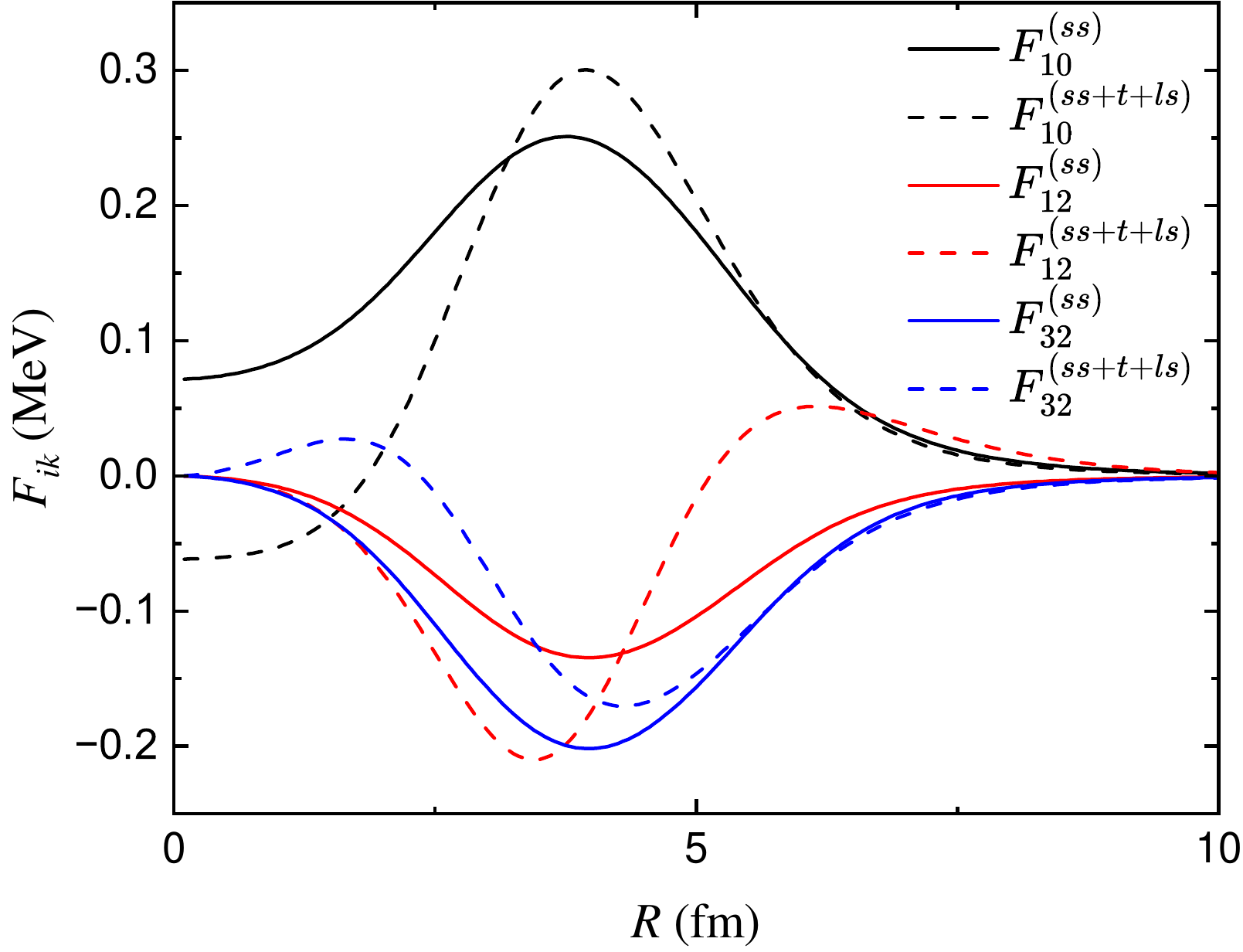}
\caption{\label{fig:formfactors} (Color online) {The radial form factors at $E_{\rm lab}=30$~MeV} for the $n+{}^{93}$Nb system. The solid curves show the spin-spin interaction contributions by the direct and exchange central spin-spin pieces. In the present DWBA work, the results are recalculated at each energy point. The dashed curves are obtained by adding the tensor and spin-orbit parts to the central spin-spin contribution as presented in Eqs.~(\ref{eq:fik_reduced_f10})--(\ref{eq:fik_reduced_f32}).}
\end{figure}

Figure~\ref{fig:formfactors} shows representative radial form factors for the
{$n+{}^{93}\mathrm{Nb}$ system at $E_{\rm lab}=30$~MeV.} The central spin-spin
 form factors, consisting of $F_{10}^{(ss)}$, $F_{12}^{(ss)}$,
and $F_{32}^{(ss)}$, are used in the neutron scattering benchmark, whereas the
form-factor set including the tensor and spin-orbit terms
illustrates the effect of the additional spin-dependent pieces. 
 This comparison tests whether adding the tensor
and spin-orbit terms can remove the residual structure. These terms modify the
magnitude and radial structure of the form factors, but they do not remove the localized $^{93}$Nb residual, as discussed in Sec.~\ref{sec:neutron-observables}.

\subsection{Comparison with the spin observables in neutron scattering}
\label{sec:neutron-observables}

\begin{figure}[!t] 
  \centering
  \includegraphics[width=0.48\textwidth,keepaspectratio]{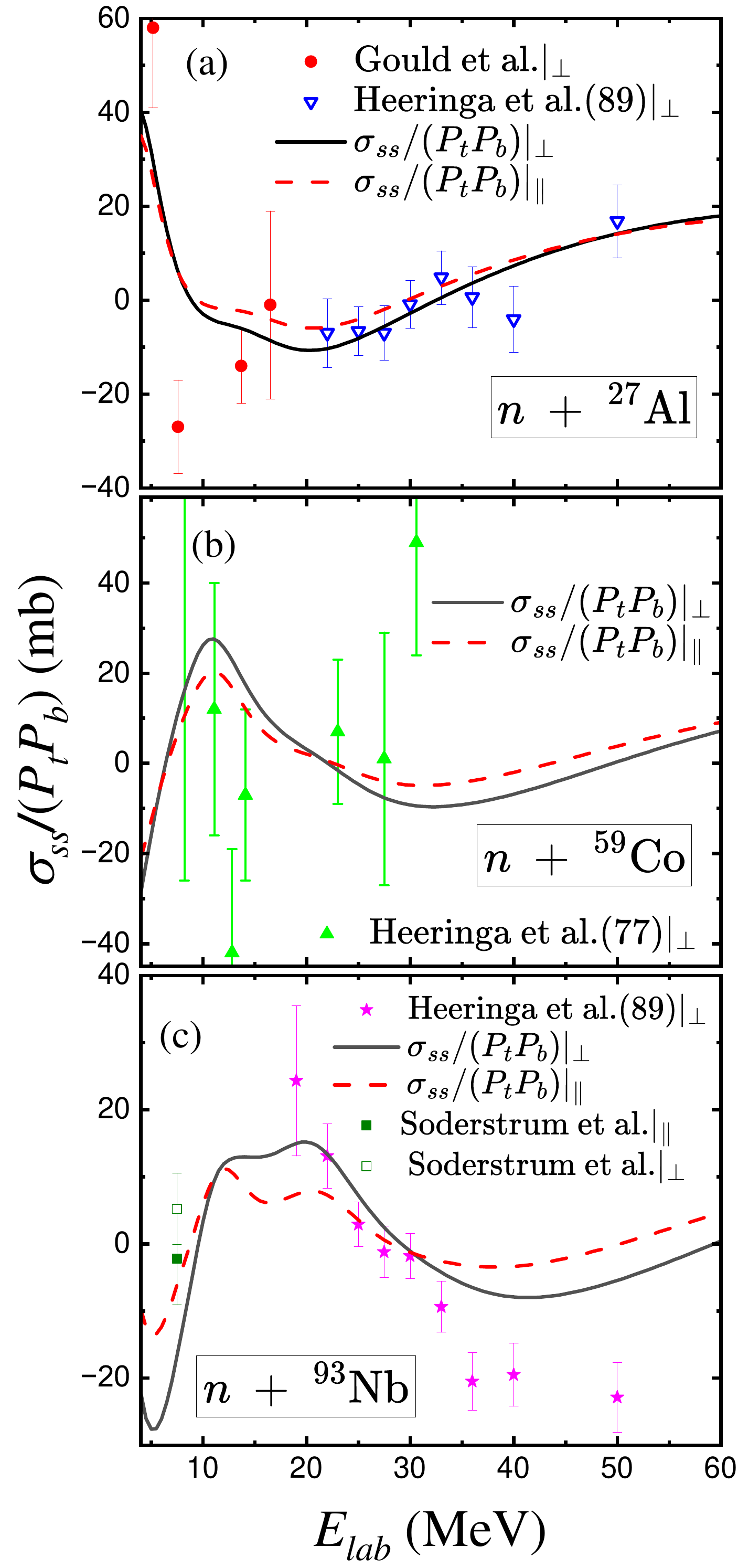}
\caption{\label{fig:benchmark-ss} (Color online) The perpendicular (solid curves) and 
the parallel (dashed curves) spin-spin observables for {neutron scattering off} 
$^{27}$Al, $^{59}$Co, and $^{93}$Nb. The results are obtained by the DWBA calculation and the various symbols are the experimental data taken from Refs.~\cite{Gould1986,Heeringa1977,Heeringa1989,Soderstrum1992}. The spin-orientation factors are listed in Table~\ref{tab:orientation_factors}.}
\end{figure}

Fig.~\ref{fig:benchmark-ss} presents the comparison between the calculated and experimental spin-spin observables for 
{neutron scattering off} $^{27}$Al, $^{59}$Co, and $^{93}$Nb. The calculations reproduce the main sign systematics and the overall scale of the measured observables for the three nuclei, while a localized low-energy residual remains most visibly in the $n+{}^{93}$Nb system. For all of the three nuclei, the folding potentials are recalculated at each energy, so that the energy-dependent exchange spin-spin interaction and the DWBA propagation use the same neutron energy. At each energy, the calculation includes the three central spin-spin form factors, $F^{(ss)}_{10}$, $F^{(ss)}_{12}$, and $F^{(ss)}_{32}$, shown in Fig.~\ref{fig:formfactors}.

Although Ref.~\cite{McAbee1990} adopts the perpendicular geometry, we show the results for both the parallel and the perpendicular geometries. 
The orientation factors entering this assembly are given in Table~\ref{tab:orientation_factors}. These tabulated input data specify the thermal substate distribution for each target. 
{{This qualitative agreement provides a check on the sign and scale of the calculated neutron observables.} It does not independently establish the absolute normalization or the uniqueness of the interaction.}

As a numerical check, we repeated the neutron scattering DWBA calculation with and without the
conventional projectile spin-orbit term in the distorting
{potential. Over $E_{\rm lab}=4$--$60$~MeV, the largest change in the perpendicular observable is about $4.74$~mb for $^{93}$Nb at $12$~MeV with central spin-spin transition form factors. The sign systematics and the localized $^{93}$Nb residual remain. The baseline curves use central spin-spin transition form factors and include the conventional projectile spin-orbit term in the distorting potential.}

The remaining low-energy residual in the $^{93}$Nb system is localized and, within the
present calculation, is found to be associated mainly with the $l=6$ contribution to the
parallel rank-0 DWBA component $\sigma_{10}$. The partial-wave analysis and the
imaginary diagnostic are discussed in Sec.~\ref{sec:discussion}.

The neutron scattering results lead to two important conclusions. First, the sign and overall {scale of the observables} are consistent with the data. {Second, the remaining discrepancies are not removed when the tensor and
spin-orbit contributions are added to the calculation including only the
central spin-spin components of $F_{10}$, $F_{12}$, and $F_{32}$.}

This distinction is essential for the subsequent application of the formalism 
to fusion in $p+{}^{93}$Nb. Since the dominant central spin-spin form factors and the operator conventions 
 have been checked against neutron observables, the fusion calculation provides a controlled estimate of the expected scale of the corresponding like-channel correction.


\section{Impact on Near-Barrier Proton Fusion}
\label{sec:proton}

Using the conventions established in the DWBA analysis of neutron scattering, we now examine the dynamical impact of the reconstructed interaction in the like-nucleon channel on proton-induced fusion near the Coulomb barrier. 

\subsection{Coupled-channels framework}

The fusion cross sections are computed 
for the $p+{}^{93}\mathrm{Nb}$ system within a coupled-channels framework. The interaction is reconstructed in the like-nucleon channel using the same operator decomposition introduced in Sec.~\ref{sec:framework}. The channel basis is defined as
\begin{equation}
|\alpha;J\rangle = |[(I_t s_p)S_\alpha, l_\alpha]J\rangle. 
\end{equation}
Here $I_t$ and $s_p$ denote the spin of the target and the incident proton, respectively. They are coupled to the channel spin $S_\alpha$, which is then coupled to the relative orbital angular momentum $l_\alpha$ to form the total angular momentum $J$.
The corresponding coupled radial equations for fixed total angular momentum $J$ and parity are given as

\begin{equation}
\left[
-\frac{\hbar^2}{2\mu}\frac{d^2}{dR^2}
- E
\right]
\chi_{\alpha}^{J}(R)
+
\sum_{\beta}
U_{\alpha\beta}(R)\,
\chi_{\beta}^{J}(R)
=0,
\label{eq:cc_radial}
\end{equation}
where $\chi_{\alpha}^{J}(R)$ is the radial wave function in channel $\alpha$ and $\mu$ is the proton-target reduced mass. The diagonal potential is written as
\begin{equation}
U_{\alpha\beta}^{(0)}(R)
=
\left[
V_N(R)+V_C(R)
+
\frac{\hbar^2 l_\alpha(l_\alpha+1)}{2\mu R^2}
\right]\delta_{\alpha,\beta},
\label{eq:diagonal_reference_potential}
\end{equation}
where $V_N$ and $V_C$ are the real nuclear and Coulomb potentials,
respectively, and the centrifugal term is therefore channel dependent through
$l_\alpha$. 
The full coupled potential is then given by
\begin{equation}
U_{\alpha\beta}(R)
=
U_{\alpha\beta}^{(0)}(R)
+
U^{(\mathrm{spin+tens+ls})}_{\alpha\beta}(R),
\label{eq:full_channel_potential}
\end{equation}
  The spin-dependent part is constructed from the form factors
$F_{10}$, $F_{12}$, and $F_{32}$.
The spin-dependent coupling matrix is evaluated with the same sign
convention as in Eq.~(\ref{eq:interaction_decomposition}),

\begin{equation}
U_{\alpha\beta}^{(\mathrm{spin+tens+ls})}(R)
=
-\sum_{(i,k)\in\mathcal S}
F_{ik}(R)
\left\langle \alpha;J \left|S_{ik}\right| \beta;J \right\rangle ,
\label{eq:spin_channel_matrix}
\end{equation}
where the retained operator set is $\mathcal S=\{(1,0),(1,2),(3,2)\}$.
{Tensor contributions are included in $F_{10}$, $F_{12}$, and $F_{32}$. For $p+{}^{93}\mathrm{Nb}$, the direct NN spin-orbit contribution is retained in $F_{10}$; the spin-orbit contributions to $F_{12}$ and $F_{32}$ are omitted.}

The channel Hamiltonian is constructed from the reference potential given by
the real central component of the proton optical potential and the adopted
$F_{10}$, $F_{12}$, and $F_{32}$ radial form factors. With this channel Hamiltonian, the coupled radial
Schr\"odinger equations are solved numerically. For the proton application, we
use the real central component of the global proton optical potential for the
$p+{}^{93}\mathrm{Nb}$ system, with the geometry parameters taken from
Table~VI of Ref.~\cite{Varner1991}, recalculated at each energy point.
Since the proton near-barrier energies considered here lie below the main
energy interval used in the global proton fit, this real central component
is used as a fixed reference baseline rather than as an optimized
$p+{}^{93}\mathrm{Nb}$ elastic-scattering or fusion potential; the proton-fusion calculation therefore quantifies the relative scale of the reconstructed spin-dependent correction on this baseline.
To isolate the role of the spin-dependent interaction, three levels of approximation are considered:
\begin{itemize}
\item \textit{Central}: spin-dependent terms are omitted,
\item \textit{Diagonal}: only the diagonal matrix elements of
$U_{\alpha\beta}^{(\mathrm{spin+tens+ls})}$ defined in
Eq.~(\ref{eq:spin_channel_matrix}) are retained,
\item \textit{Full CC}: the complete coupling matrix is included.
\end{itemize}

The total fusion cross section is obtained from the entrance-channel
penetrabilities as
\begin{align}
\sigma_{\mathrm{fus}}(E)
&=
\frac{\pi}{k^2}
\frac{1}{(2I_t+1)(2s_p+1)}
\nonumber\\
&\quad\times
\sum_{\pi,J}
(2J+1)
\sum_{\alpha\in\mathrm{entrance}}
T_{\alpha}^{J\pi}(E)~,
\label{eq:fusion_cross_section}
\end{align}
where $k$ is the incident wave number 
in the center-of-mass frame and $T_{\alpha}^{J\pi}$ is the penetrability for entrance channel
$\alpha$ in a fixed-$J$ and fixed-parity block. The calculation sums even- and odd-parity sectors up to $l_{\max}=7$ and $J=11$. 

The coupled radial equations are solved with the Incoming Wave Boundary Condition (IWBC) and modified-Numerov
scheme of CCFULL~\cite{Hagino1999}, using the fixed-$(J,S,l)$ channel
Hamiltonian defined above rather than the standard deformation-channel input.
That is, fusion penetrabilities are obtained with IWBC at the pocket of the
entrance reference potential. The coupled equations are propagated with a
modified Numerov algorithm and matched to Coulomb asymptotic functions at
$R_{\max}=40$~fm using the mesh size of $0.05$~fm.

In the proton-fusion calculations reported below, no
additional imaginary optical potential is introduced; absorption into fusion is
represented by the incoming-wave boundary condition, while the imaginary
$F_{10}$ term discussed in Sec.~\ref{sec:discussion} is used only as a separate
phenomenological diagnostic.

\subsection{Radial structure and barrier properties}

In the channel basis, two entrance spin sectors are the $S=4$ and $S=5$ diagonal sectors {by $I_t = 9/2 \otimes s_p (= 1/2) \rightarrow S = 4,5$}. The largest off-diagonal couplings occur within the $J=4$ and $J=5$ even-parity subspaces. Representative channels are $J=4$, $S=4$, $l=0\leftrightarrow2$ and $J=5$, $S=5$, $l=0\leftrightarrow2$: 
{the off-diagonal potentials for these channels reach about $+0.338$ and $-0.257$~MeV at $R\approx4.60$ and $4.30$~fm, respectively, but at the barrier position, $R_B=8.62$~fm, they are reduced to about $+2.15$ and $+3.10$~keV, respectively. This is the relevant scale in the present fusion reaction, since the penetration in the fusion is governed primarily by the barrier region rather than by the interaction strength in the nuclear interior.}

\begin{figure*}[!t] 
  \centering
  \includegraphics[width=0.92\textwidth,height=0.55\textheight,keepaspectratio]{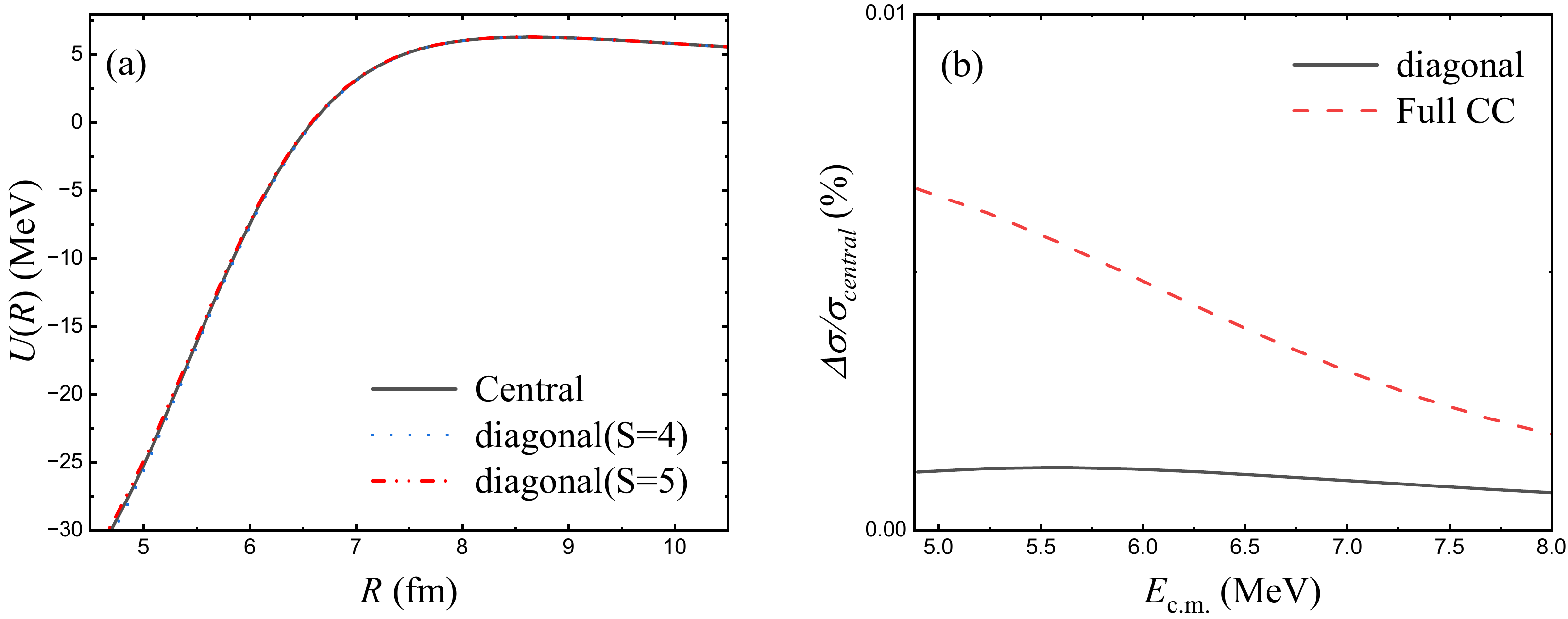}
\caption{\label{fig:barrier-fusion} (Color online)
Analysis of the $p+{}^{93}$Nb fusion reaction using the
interaction in the like-nucleon channel reconstructed with the conventions
established in the DWBA analysis of neutron scattering and the reference
potential generated by the real central component of the proton optical
potential, with the
geometry parameters taken from Table~VI in Ref.~\cite{Varner1991}. Panel~(a) shows the central-plus-Coulomb barrier together with the diagonal potentials obtained after adding the reconstructed spin-dependent correction in the $S=4$ and $S=5$ entrance-channel spin sectors at the near-barrier reference energy $E_{\rm lab}=6.363$~MeV (see also Table~\ref{tab:proton-scales}).
Panel~(b) shows the relative changes of the fusion cross sections,
$(\sigma_{\rm variant}/\sigma_{\rm central})-1$, obtained from the
energy-dependent proton fusion calculation with the fixed-$(J,S,l)$
coupled-channels solver described in Sec.~\ref{sec:proton}. 
The {relative changes} in panel~(b) and Table~\ref{tab:fusion} are obtained from energy-matched calculations at each energy.
In panel~(b), ``diagonal'' denotes the unpolarized fusion result obtained after retaining only the diagonal spin-dependent matrix elements in the $S=4$ and $S=5$ entrance-channel spin sectors; it is not an $S$-resolved cross section.
The real spin-dependent correction changes the {unpolarized} proton fusion cross section only {slightly within} the present {model}.}
\end{figure*}

Figure~\ref{fig:barrier-fusion} summarizes the proton-fusion result for the
$p+{}^{93}\mathrm{Nb}$ system using the like-channel $F_{10}$, $F_{12}$,
and $F_{32}$ form factors reconstructed with the conventions established in the DWBA analysis of neutron scattering. The proton-fusion calculation is performed energy by energy: at each fusion-energy point, both the real central component of the CH89 proton optical potential constructed with the geometry parameters of Table~VI in Ref.~\cite{Varner1991} and the like-channel $F_{10}$, $F_{12}$, and $F_{32}$ form factors are recalculated at the corresponding $E_{\rm lab}$ and then used in the fixed-$(J,S,l)$ coupled-channels calculation described above.

{At the near-barrier reference energy $E_{\rm lab}=6.363$~MeV, the spin-dependent correction splits the diagonal entrance-channel potentials for $S=4$ and $S=5$ in opposite directions, but the shifts remain below $10$~keV at the central barrier. Here the central barrier is $R_B=8.62$~fm and $\quad{V_B=6.295}$~MeV. The $S=4$ barrier is lowered and the $S=5$ barrier is raised.}

Table~\ref{tab:proton-scales} collects representative scales in the barrier region for the near-barrier reference calculation at $E_{\rm lab}=6.363$~MeV used in panel~(a) of Fig.~\ref{fig:barrier-fusion}. These numbers make it clear why the proton effect remains tiny: the listed $F_{10}$ form factor is already only of order $10^{-3}$--$10^{-2}$~MeV at the barrier radius, the diagonal barrier shifts stay below $10$~keV, and the surviving off-diagonal couplings at $R_B$ are likewise only a few $10^{-3}$~MeV.

A further suppression arises from the unpolarized entrance-spin average.
For $I_t=9/2$ and $s_p=1/2$, the two entrance spin sectors $S=4$ and $S=5$
carry statistical weights $9/20$ and $11/20$, respectively. The diagonal
spin-dependent barrier shifts in Table~\ref{tab:proton-scales} have opposite signs and
nearly cancel under this average,
$(9/20)\Delta V_B(S=4)+(11/20)\Delta V_B(S=5)
\simeq {2.4}\times10^{-5}~\mathrm{MeV}$. Thus the diagonal barrier splitting is
largely traceless under the unpolarized spin average. The residual fusion
modification is therefore controlled by the small nonlinearity of the
penetrability, the weak off-diagonal couplings at $R_B$, and the radial
localization of the spin-dependent form factors.

\begin{table}[t] 
\caption{\label{tab:proton-scales}Representative scales in the barrier region at $E_{\rm lab}=6.363$~MeV, corresponding to the panel~(a) of Fig.~\ref{fig:barrier-fusion}. The diagonal shifts are evaluated relative to the central barrier, and the two off-diagonal entries correspond to the $J=4$, $S=4$, $l=0\leftrightarrow2$ and $J=5$, $S=5$, $l=0\leftrightarrow2$ couplings at the same barrier radius. Here ``w/o LS'', ``LS'', and ``with LS'' denote the reconstructed \(F_{10}\) without the spin-orbit term, the spin-orbit contribution itself, and the result including the spin-orbit term, respectively. {{The quoted precision reflects the numerical calculation and does not represent an uncertainty estimate for the model.}}}
\begin{ruledtabular}
\begin{tabular}{lc}
Quantity & Value \\
\hline
$R_B$ & {$8.62~\mathrm{fm}$} \\
$V_B$ & {$6.295~\mathrm{MeV}$} \\
$F_{10}^{\rm w/o\,LS}(R_B)$ & {$-5.09~\mathrm{keV}$} \\
$F_{10}^{\rm LS}(R_B)$ & {$-0.139~\mathrm{keV}$} \\
$F_{10}^{\rm with\,LS}(R_B)$ & {$-5.23~\mathrm{keV}$} \\
$\Delta V_B(S=4)$ & {$-6.357~\mathrm{keV}$} \\
$\Delta V_B(S=5)$ & {$+5.245~\mathrm{keV}$} \\
$V_{J=4,S=4,l=0\leftrightarrow2}(R_B)$ & {$+2.15~\mathrm{keV}$} \\
$V_{J=5,S=5,l=0\leftrightarrow2}(R_B)$ & {$+3.10~\mathrm{keV}$} \\
\end{tabular}
\end{ruledtabular}
\end{table}

\subsection{Fusion cross sections}

Figure~\ref{fig:barrier-fusion}(b) shows the resulting relative change in the fusion cross section. Since the central, diagonal-only, and full coupled-channels results are nearly indistinguishable on the absolute scale, we plot the relative ratio given by 
\begin{equation}
\frac{\Delta \sigma}{\sigma_{\mathrm{central}}}
=
\frac{\sigma_{\mathrm{variant}}}{\sigma_{\mathrm{central}}} - 1,
\end{equation}
where $\sigma_{\mathrm{variant}}$ denotes either the diagonal-only or the full coupled-channels result.

Figure~\ref{fig:barrier-fusion}(b) and Table~\ref{tab:fusion} show that
the full parity-summed coupled-channels correction is systematic but extremely
small. {At the energies listed in Table~\ref{tab:fusion}, the full-CC result differs from the central case by about $(3$--$5)\times10^{-5}$ in ratio, or $0.003$--$0.005$~\%.} This {range indicates the scale within the adopted model, not its physical accuracy. This} reflects the weak coupling strength of
the spin-dependent interaction in the barrier region.
{Across the eleven calculated energies in the range $E_{\rm cm}\approx4.90$--$8.40$~MeV, the absolute full-CC correction remains below $0.01$~\%, with its largest value at the lowest energy.}
In practical terms, the present calculation does not indicate that spin dependence is irrelevant in general, but suggests that the present real interaction produces only a very weak effect in this specific proton application. Related extended-optical-model studies of weakly bound systems, including the corresponding erratum for the $^{17}$F breakup study, likewise show that elastic, breakup, and fusion observables must be interpreted together when one assesses near-barrier coupling effects~\cite{Heo2024Ne17,Heo2022F17DPP,Heo2024F17Erratum}.

\begin{table}[t] 
\caption{\label{tab:fusion}Representative near-barrier parity-summed coupled-channels fusion cross sections {and relative corrections}. For each row, both the real central component of the global proton optical potential and the like-channel spin-dependent form factors are recalculated with the Table-VI geometry parameters of Ref.~\cite{Varner1991} at the corresponding fusion-energy point through the same $E_{\rm lab}$ and then used in the fixed-$(J,S,l)$ coupled-channels calculation described above.
The calculation sums even- and odd-parity sectors up to $l_{\max}=7$ and $J=11$. The first column lists $E_{\mathrm{cm}}-V_B(E)$, where $V_B(E)$ is the barrier height of the recalculated central baseline at that energy{. The} relative {corrections are rounded} to {one significant digit to indicate their scale within} the {present model}.}
\centering
\renewcommand{\arraystretch}{1.1}
\setlength{\tabcolsep}{5pt}
\resizebox{\columnwidth}{!}{\begin{tabular}{@{}ccccc@{}}
\toprule
$E_{\mathrm{cm}}-V_B(E)$ & Central& $\frac{\sigma_{\rm diag}}{\sigma_{\rm central}}-1$ & $\frac{\sigma_{\rm Full\,CC}}{\sigma_{\rm central}}-1$ & Regime \\
(MeV) & (mb) & & &  \\
\midrule
$-0.35$ & {$197$} & {$1\times10^{-5}$} & {$5\times10^{-5}$} & Below barrier \\
$+0.00$ & {$255$} & {$1\times10^{-5}$} & {$4\times10^{-5}$} & At barrier \\
$+0.35$ & {$316$} & {$1\times10^{-5}$} & {$4\times10^{-5}$} & Above barrier \\
$+1.05$ & {$441$} & {$9\times10^{-6}$} & {$3\times10^{-5}$} & Above barrier \\
\bottomrule
\end{tabular}}
\end{table}

\subsection{Physical interpretation}

The calculation for proton fusion shows that, after the dominant central spin-spin form factors and operator conventions are checked against neutron-scattering observables, the corresponding like-channel correction changes the fusion cross section only weakly. The suppression comes from two related effects: First, the spin-dependent form factors are weak in the barrier region. Second, the spin-dependent interaction produces barrier shifts of opposite sign in the diagonal $S=4$ and $S=5$ channels and largely cancels after the unpolarized spin average.

In the present system, the spin-dependent terms are localized at radii smaller than the barrier radius and therefore have limited influence on the tunneling probability. In contrast, neutron observables probe the full radial region through distorted-wave propagation, which explains their stronger sensitivity to the corresponding spin-dependent operator structure.

The present calculation therefore indicates that {the retained real} spin-dependent interaction has only a limited influence on the {unpolarized} near-barrier {proton} fusion observables considered here.

\subsection{Extension to a lighter system: \texorpdfstring{$p + {}^{27}\mathrm{Al}$}{p + 27Al}}

\begin{figure*}[!t] 
  \centering
  \includegraphics[width=0.92\textwidth,height=0.55\textheight,keepaspectratio]{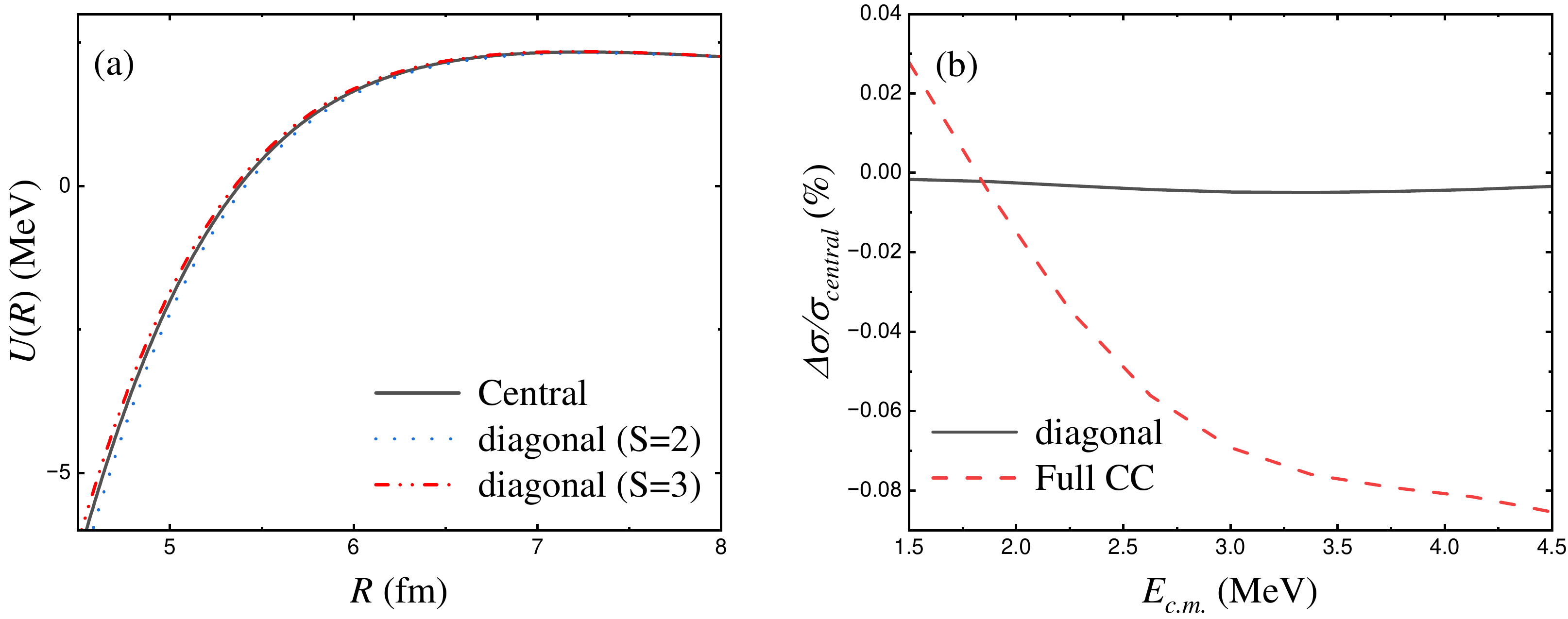}
\caption{\label{fig:barrier-fusion-al27} (Color online)
Same as Fig.~\ref{fig:barrier-fusion}, but for the 
$p+{}^{27}\mathrm{Al}$ system. For this system, the corresponding entrance-channel spin sectors are $S=2$ and $S=3$. {Both panels use the central spin-spin and direct tensor contributions, with the direct NN spin-orbit terms omitted.}}
\end{figure*}

As a further comparison, we also perform 
the calculation for the lighter system $p+{}^{27}\mathrm{Al}$. 
 The result is shown in
{Fig.~\ref{fig:barrier-fusion-al27}, which {also shows a small} near-barrier correction for the present real interaction. {The absolute full-CC correction remains below $0.1$~\% over the nine calculated energies $E_{\rm cm}=1.5$--$4.5$~MeV. Its largest value corresponds to a reduction of the fusion cross section at $E_{\rm cm}=4.5$~MeV.}}

Table~\ref{tab:p27al_scales} summarizes the representative numerical scales
 shown in
Fig.~\ref{fig:barrier-fusion-al27}. The barrier quantities are quoted for the
reference $E_{\rm lab}=2.424$~MeV calculation used in panel~(a), while the
maximum changes of the fusion ratio are taken over the energy range shown in
{panel~(b). The sign change of the full-CC curve in Fig.~4(b) indicates that the detailed interference pattern is system dependent.}

\begin{table}[!h]
\caption{\label{tab:p27al_scales}Representative near-barrier scales for the $p+{}^{27}$Al system. 
 The notation of \(F_{10}\) is the same as in Table~\ref{tab:proton-scales}. {{The barrier shifts and changes of the fusion-ratio are calculated without the direct NN spin-orbit terms.} The LS and with-LS values of $F_{10}$ are listed only for comparison and are not used in Fig.~\ref{fig:barrier-fusion-al27}.}
}
\begin{ruledtabular}
\begin{tabular}{lc}
Quantity & Value \\
\hline
$R_B$ & {$7.24~\mathrm{fm}$} \\
$V_B$ & {$2.337~\mathrm{MeV}$} \\
$F_{10}^{\rm no\,LS}(R_B)$ & {$-9.61~\mathrm{keV}$} \\
$F_{10}^{\rm LS}(R_B)$ & {$-0.122~\mathrm{keV}$} \\
$F_{10}^{\rm with\,LS}(R_B)$ & {$-9.73~\mathrm{keV}$} \\
$\Delta V_B(S=2)$ & {$-13.21~\mathrm{keV}$} \\
$\Delta V_B(S=3)$ & {$+9.745~\mathrm{keV}$} \\
$\max|\sigma_{\rm diag}/\sigma_{\rm central}-1|$ & {$5\times10^{-5}$} \\
$\max|\sigma_{\rm full~CC}/\sigma_{\rm central}-1|$ & {$9\times10^{-4}$} \\
\end{tabular}
\end{ruledtabular}
\end{table}

\section{Discussion}
\label{sec:discussion}

The remaining discrepancy in the neutron observables of the $n+{}^{93}\mathrm{Nb}$ system shown in Fig.~\ref{fig:benchmark-ss} provides important insight into the limitations of the present framework. {{Adding the tensor and NN spin-orbit contributions, or including projectile spin-orbit distortion, does not remove the remaining discrepancy.}} In that respect it is also consistent with the broader polarized-neutron literature, where spin observables can be sensitive to reaction-mechanism and absorptive details even when the central spin-dependent signal is qualitatively understood~\cite{Goodman1985,HnizdoKemper1987,Nagadi2004}.

{A partial-wave decomposition shows a strong $l=6$ contribution to the low-energy structure. The parallel component $\sigma_{10}(l=6)$ is about $+2.26$, $-5.16$, and $-5.72$~mb at the representative energies $E_{\rm lab}=12.5$, $16.0$, and $20.5$~MeV, respectively. Removing only the incoming-$l=6$ contribution of $F_{10}$ from the full forward amplitude raises the perpendicular observable from $13.0$ to $20.0$~mb at $16.0$~MeV and from $15.0$ to $23.4$~mb at $20.5$~MeV. Table~\ref{tab:l6-residual} gives representative values.}

\begin{table}[t] 
\caption{\label{tab:l6-residual} Partial cross sections of $\sigma_{10}$ in the parallel geometry with and without the $l=6$ contribution at representative low-energy points in the $n+{}^{93}$Nb double-peak region. {The last two columns show the perpendicular observable with and without the incoming-$l=6$ contribution of $F_{10}$, while the $F_{12}$ and $F_{32}$ responses are unchanged.}}
\begin{ruledtabular}
\begin{tabular}{cccccc}
$E_{\rm lab}$ & $\sigma_{10}$ & $\sigma_{10}(l=6)$ & $\sigma_{10}(l\neq6)$ & $\left.\frac{\sigma_{ss}}{P_tP_b}\right|_{\perp}$ & $\left.\frac{\sigma_{ss}}{P_tP_b}\right|_{\perp,\;l\neq6}$  \\
(MeV) & (mb) & (mb) & (mb) & (mb) & (mb) \\
\hline
12.5 & {$+8.65$} & {$+2.26$} & {$+6.39$} & {$+12.5$} & {$+10.4$} \\
16.0 & {$+6.27$} & {$-5.16$} & {$+11.43$} & {$+13.0$} & {$+20.0$} \\
20.5 & {$+7.90$} & {$-5.72$} & {$+13.62$} & {$+15.0$} & {$+23.4$} \\
\end{tabular}
\end{ruledtabular}
\end{table}

\begin{figure*}[t] 
  \centering
  \includegraphics[width=0.96\textwidth,height=0.40\textheight,keepaspectratio]{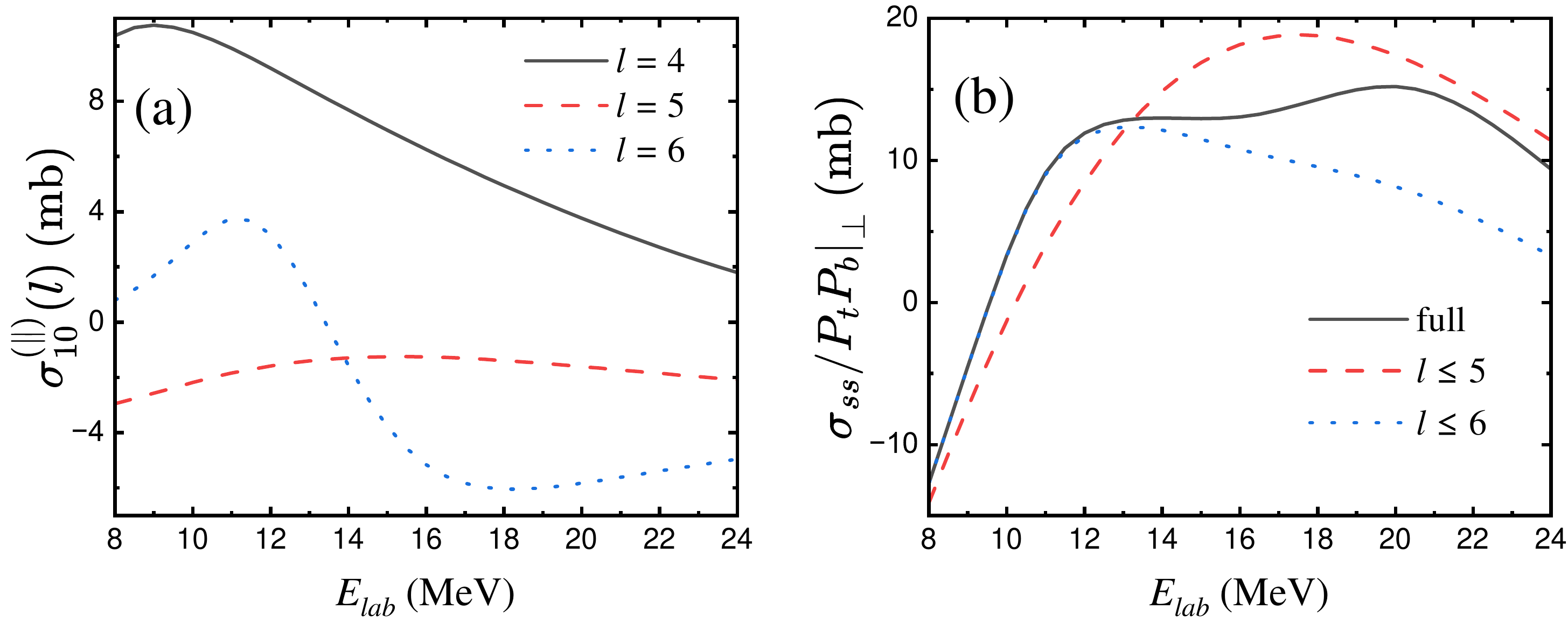}
  \caption{\label{fig:l6-contrast} (Color online) 
   Comparison of the contribution of the $l=6$ partial wave with the contributions of the neighboring low-$l$ partial waves in the $n+{}^{93}$Nb system. The left panel shows the contributions of the $l=4$, $l=5$, and $l=6$ partial waves to the parallel component $\sigma_{10}$. The $l=4$ and $l=5$ terms vary comparatively smoothly, {whereas the $l=6$ term changes sign between $13.0$ and $13.5$~MeV and becomes the dominant negative contribution around $16$--$21$~MeV.} The right panel shows the perpendicular observable reconstructed from cumulative partial-wave sums. {The $l\leq6$ sum has one maximum, whereas the full curve has two on the calculated energy grid; higher partial waves are therefore needed for the second maximum.}}
\end{figure*}

Figure~\ref{fig:l6-contrast} shows the contrast with the neighboring partial waves explicitly. The left panel compares the $l=4$, $l=5$, and $l=6$ contributions to $\sigma_{10}$ and shows that only the $l=6$ term changes sign strongly in the dip region. {The right panel shows that adding $l=6$ changes the low-energy maximum and suppresses the observable near the dip, but the second maximum in the full curve also requires higher partial waves. Diagnostic variants with stronger absorption suppress the magnitude of this $l=6$ interference. The result depends on absorption, but this test does not establish absorption as the unique cause of the discrepancy.}

\begin{figure*}[!t] 
  \centering
  \includegraphics[width=0.92\textwidth,height=0.55\textheight,keepaspectratio]{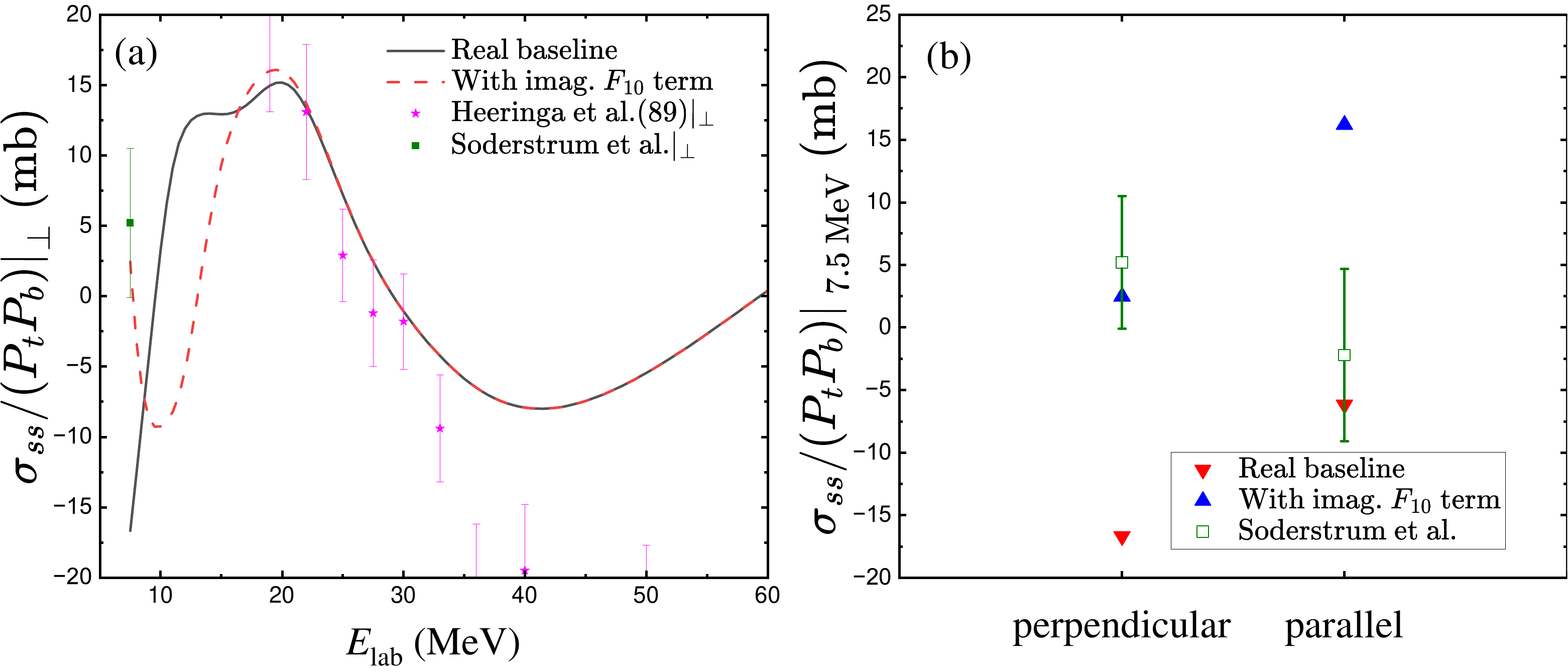}
\caption{\label{fig:imag-discussion} (Color online) Diagnostic role of the phenomenological imaginary $F_{10}$ term. 
 Panel (a) compares the real neutron baseline (`Real baseline': the black solid line) with one representative low-energy-localized imaginary-$F_{10}$ case (`with imag.\ $F_{10}$ term': the red dashed line) generated from Eq.~(\ref{eq:artificial_imag}) with the {representative diagnostic} values $\eta=+8.86$ and $E_0=9.25$~MeV on the updated energy-matched baseline consisting of the central spin-spin parts of $F_{10}$, $F_{12}$, and $F_{32}$ for $^{93}$Nb. The energy-dependent transverse data are from Heeringa et al.~\cite{Heeringa1989}, and the additional $7.5$~MeV transverse point is from Soderstrum et al.~\cite{Soderstrum1992}. Panel (b) isolates the transverse and 
the longitudinal contributions at $7.5$~MeV in comparison with the Soderstrum et al.~\cite{Soderstrum1992} data. The imaginary term can move the perpendicular/transverse point towards the experimental value, but it simultaneously drives the parallel/longitudinal point to a large positive value inconsistent with the measured constraint. Accordingly, the imaginary term is retained only as a diagnostic and is not used in the main neutron or proton baseline curves.}
\end{figure*}

We nevertheless examined a phenomenological imaginary diagnostic term as a probe of missing spin-dependent absorption. This step was motivated by the long-standing observation that real and imaginary spin-dependent terms can drive qualitatively different energy dependences in neutron spin observables~\cite{Gould1986,Heeringa1977,HusseinSherif1973,Nagadi2004}, together with a modern optical-model study, including its corresponding erratum, in which dynamic-polarization and absorptive effects must be tracked explicitly~\cite{Heo2022F17DPP,Heo2024F17Erratum}. In its simplest form the test modifies the $F_{10}$ component as
\begin{equation}
F_{10}(r;E)
\rightarrow
F_{10}(r;E)
+
i\,\eta\,\exp\!\left[-\left(\frac{E}{E_0}\right)^2\right]F_{10}(r;E),
\label{eq:artificial_imag}
\end{equation}
where $\eta$ is a dimensionless strength parameter and $E_0$ sets the energy scale of the Gaussian damping factor. {In this diagnostic, $E=E_{\rm lab}$.} 
Both parameters are used only as phenomenological diagnostics. Four phenomenological forms were examined:
\begin{enumerate}
\item energy-independent imaginary terms applied to all three radial form factors, $F_{10}$, $F_{12}$, and $F_{32}$,
\item energy-independent imaginary terms applied to $F_{10}$ only,
\item low-energy-localized terms of the form $\exp[-(E/E_0)^2]$ applied to all three retained form factors,
\item the same low-energy-localized form applied only to $F_{10}$.
\end{enumerate}
The broad scan used $\eta\in[-3,3]$ and a reference low-energy weight $\exp[-(E/15\,\mathrm{MeV})^2]$, while the refined $F_{10}$-only scan used $\eta\in[-10,10]$ and $E_0\in[5,25]$~MeV. The representative neutron comparison shown in Fig.~\ref{fig:imag-discussion} uses the fourth ansatz above, namely the low-energy-localized term applied only to $F_{10}$, with $\eta = +8.86$ and $E_0 = 9.25$ MeV {as representative diagnostic values}.

Panel~(a) of Fig.~\ref{fig:imag-discussion} compares the perpendicular (transverse) observable with the energy-dependent data of Heeringa et al.~\cite{Heeringa1989} together with the additional $7.5$~MeV transverse point of Soderstrum et al.~\cite{Soderstrum1992}, while panel~(b) of the same figure isolates the perpendicular (transverse) and the parallel (longitudinal) constraints at 7.5 MeV from Soderstrum et al.~\cite{Soderstrum1992}. This comparison highlights the central limitation of the phenomenology: the transverse point at 7.5 MeV can be brought closer to the experimental value only at the cost of driving the longitudinal point to a large positive value inconsistent with the measured constraint. A further combined scan allowing a common real rescaling $\lambda_R\in[0.5,1.5]$ of the real baseline, $F_{ik}\rightarrow\lambda_R F_{ik}$, together with $\eta\in[-12,12]$ did not remove this tension; {with $E_0=9.25$~MeV fixed, the best fit was obtained near $\lambda_R=1.325$ and $\eta=+8.02$.}

The phenomenological form of Eq.~(\ref{eq:artificial_imag}) can also be applied to the proton fusion calculation as a diagnostic, although that proton response is not plotted in Fig.~\ref{fig:imag-discussion}. In that separate test we apply the same $(\eta,E_0)=(8.86,9.25~\mathrm{MeV})$ ansatz to the proton $F_{10}$ component while keeping the spin-orbit term real. {This calculation is used only to examine the sensitivity to an}
imaginary $F_{10}$ component and is not included in the retained real fusion
baseline; we therefore do not interpret the proton response quantitatively
here. This diagnostic term is not intended as a fitted spin-dependent optical potential. 

  The radial peaks of the reconstructed spin-dependent form factors lie well inside the barrier radius, so {their effect on near-barrier penetration is limited}. More spatially extended valence densities could in principle increase the overlap with the barrier region, which is why weakly bound or halo-like systems remain interesting future targets for the same framework~\cite{Canto2006,HaginoTakigawa2012,Heo2020Be11DPP,Heo2024Ne17}. Extended optical-model analyses of halo and weakly bound systems provide concrete examples in which long-range polarization and breakup dynamics become essential once the valence density reaches the barrier region~\cite{Heo2020Be11DPP,Heo2022F17DPP,Heo2024Ne17}. At the same time, those systems often involve low-$l$ valence configurations and stronger coupling to breakup or transfer channels, so the balance of operators need not follow the present $^{93}$Nb case. A realistic study of that regime would therefore require additional calculations, even though it is beyond the scope of the present study.

\section{Conclusions}
\label{sec:conclusions}

{We used the neutron spin observables to examine the sign and scale of the dominant central spin-spin contribution in a nucleon-nucleus potential, with explicit operator conventions for the reconstruction in the like-nucleon channel.} The corresponding like-channel interaction was then applied to fusion in the $p+{}^{93}$Nb system without any additional adjustment of the spin-dependent terms. Thus, the fusion result provides an estimate of the expected scale of the reconstructed interaction, rather than a separate validation of the interaction in the proton channel.

   {For the retained real interaction}, the barrier modification and the resulting change in the {unpolarized} fusion cross section remain {very small in the present $p+{}^{93}$Nb calculation. The} comparison with the lighter system $p+{}^{27}$Al gives a {small correction within the adopted real interaction}. {The suppression arises from the weak spin-dependent form factors in the barrier region and the cancellation between opposite barrier shifts under the unpolarized entrance-spin average.}

   {Our conclusion is restricted} to the retained real spin-dependent interaction{. The} phenomenological imaginary $F_{10}$ term is used only as a diagnostic {and does not satisfy the transverse and longitudinal neutron constraints simultaneously. The present analysis does not provide a quantitative constraint on} spin-dependent {absorptive contributions to proton fusion.}

   {It would be interesting in future work to investigate the cancellation for suitably polarized proton beams and targets. Such configurations may reduce the cancellation associated with the unpolarized entrance-spin average and increase the sensitivity to} the spin-dependent {interaction. The form factors remain weak in the barrier region, however, and the magnitude of the effect requires a calculation with specified polarized entrance states}.

\appendix

\section{Form Factors and Observable Reduction}
\label{app:fik}

This appendix collects the equations used in the neutron scattering analysis. The purpose is not to reproduce every matrix-element reduction step of the earlier literature, but to summarize the radial quantities, assembly rules, and observable reduction.

\subsection{Direct and exchange parts in the nucleon-nucleus spin-spin interaction}
First, we obtain the direct term in the folded nucleon-nucleus spin-spin interaction, $U_k^{(ss,\mathrm{dir})}(R)$, by folding the multipole component of the projectile--valence-nucleon interaction $v_k^{(ss)} (R,\alpha r_t)$ with the valence radial density 
\begin{equation}
U_k^{(ss,\mathrm{dir})}(R)
=
\int_0^\infty dr_t\,
u_{l_t}^2(r_t)\,
v_k^{(ss)}(R,\alpha r_t),
\label{eq:uss_direct}
\end{equation}
where the direct part in the $v_k^{(ss)} (R,\alpha r_t)$ is obtained from the (unlike or like) nucleon-nucleon (NN) spin-spin radial interaction $v^{(ss)} (r_{pt})$
\begin{equation}
v_k^{(ss)}(R,\alpha r_t)
\equiv
\frac{1}{2}\int_{-1}^{1}dz\,
v^{(ss)}(r_{pt})\,P_k(z),
\label{eq:vss_multipole}
\end{equation}
where $P_k(z)$ is the Legendre polynomial of order $k$, and $\alpha$ and $r_{pt}$ are defined as
\begin{equation}
r_{pt}=\left(R^2+\alpha^2 r_t^2-2\alpha R r_t z\right)^{1/2}~,~ 
\alpha=\frac{A_t}{A_t+1},
\end{equation}
with $A_t$ being the target mass number, $r_t$ the valence radial coordinate, and $z=\cos\theta_{Rt}$ the cosine of the angle 
between $\bm R$ and $\bm r_t$. 

The exchange term is generated separately with the T4Y-based local single-nucleon-knockout exchange prescription,
\begin{equation}
U_k^{(ss,\mathrm{ex})}(R)
\equiv
U_k^{(ss,\mathrm{ex})}\!\left[k(R),k_F(R)\right],
\label{eq:uss_exchange}
\end{equation}
using the local projectile momentum $k(R)$ and the local Fermi momentum
$k_F(R)$ constructed from the same valence density $\rho_v$ used in
the direct folding term. This explicit exchange term is essential because, although the direct and exchange pieces can differ substantially in both 
{the} magnitude and {the} sign, their summed contribution is much more stable and moderate. For the central spin-spin part entering the assembled form factors, we define the direct-plus-exchange combination as
\begin{equation}
U_k^{(ss)}(R)
\equiv
-4\,U_k^{(ss,\mathrm{dir})}(R)
+4\,U_k^{(ss,\mathrm{ex})}(R).
\label{eq:Uss_codelevel}
\end{equation}

\subsection{Tensor interaction part in the nucleon-nucleus spin-spin interaction}

The tensor radial moments are built from the Horie--Sasaki 
decomposition \cite{HorieSasaki1961}, which is a tensor-multipole expansion of the
nucleon--nucleon tensor interaction in terms of two coordinates:
the external projectile--target coordinate $\bm R$ and the internal
valence coordinate $\bm r_t$.

For example, the $(k,n;2)$ channel denotes the component in which a rank-$k$
tensor in the external coordinate $\bm R$ and a rank-$n$ tensor in the
internal coordinate $\bm r_t$ are coupled to the total rank 2:

\begin{multline}
v^{(k,n;2)}(R,\alpha r_t)
\equiv
{\frac{2}{\pi}\int_0^\infty dq\,q^2
j_k(qR)j_n(q\alpha r_t)}
\\
{\times\int_0^\infty ds\,s^2v^{(t)}(s)j_2(qs)},
\label{eq:vkn2_definition}
\end{multline}
where $v^{(t)} ({s})$ is the tensor NN interaction, and {$j_\ell$ denotes a spherical Bessel function}. 
{The key idea is that the tensor force is recoupled into irreducible tensor
products of the external and internal angular variables, with total rank $2$.
This separates the angular-momentum algebra from the radial folding step.}
By folding $v^{(k,n;2)}$ 
with the same valence radial density, one obtains the folded tensor interaction
\begin{equation}
U_{kn}^{(t)}(R)
\equiv
\int_0^\infty dr_t\,
u_{l_t}^2(r_t)\,
v^{(k,n;2)}(R,\alpha r_t).
\label{eq:Ukn_definition}
\end{equation}
In this way, {the Horie--Sasaki decomposition provides the bridge from the original
tensor interaction $v^{(t)}(r_{pt})$ to the operator-basis form factors
used in the DWBA and coupled-channels calculations.}
In the Horie--Sasaki decomposition used here, the tensor radial integrals are denoted by
$U_{kn}^{(t)}(R)$, where $k$ is the external spatial rank associated with
$\bm R$ and $n$ is the internal valence-coordinate multipole rank associated
with $\bm r_t$. The tensor components needed in the present $^{93}$Nb assembly are
$(k,n)=(0,2),(2,0),(2,2),(2,4)$.

To make the connection with Eqs.~(\ref{eq:fik_reduced_f10})--(\ref{eq:fik_reduced_f32}) explicitly, we define the rescaled tensor ingredients as
$T_{kn}(R)\equiv -4U_{kn}^{(t)}(R)$. With this convention, the tensor terms in the main text correspond to the explicit components $T_{02}$, $T_{20}$, $T_{22}$, and {$T_{24}$ used below}.

The corresponding tensor contribution is given as
\begin{widetext}
\begin{align}
F_{ik}^{(t)}(R)
&= - 4
\sum_n C_{ikn}^{(t)}(I,l_t)\, U_{kn}^{(t)}(R),
\label{eq:Fik_tensor_appendix}
\\
C_{ikn}^{(t)}(I,l_t)
&=
{-}\frac{3\sqrt{5}}{2}{(2l_t+1)}
(\hat i\,\hat I\,\hat k)^2
(-1)^{(k+n)/2+l_t}
\hat n^2
\begin{Bmatrix}
2 & 1 & 1\\
i & k & n
\end{Bmatrix}
\begin{Bmatrix}
I & \tfrac12 & l_t\\
I & \tfrac12 & l_t\\
i & 1 & n
\end{Bmatrix}
\notag\\
&\quad\times
{\begin{pmatrix}
k & n & 2\\
0 & 0 & 0
\end{pmatrix}}
\begin{pmatrix}
l_t & n & l_t\\
0 & 0 & 0
\end{pmatrix}
\begin{pmatrix}
i & 1 & k\\
0 & 0 & 0
\end{pmatrix}
\begin{pmatrix}
I & I & i\\
I & -I & 0
\end{pmatrix},
\label{eq:Cikn_tensor}
\end{align}
\end{widetext}
which is the explicit coefficient formula used to connect the Horie--Sasaki single-folds to the operator basis used in the main text.

\subsection{Direct spin-orbit combinations}

The direct neutron spin-orbit contribution is built from the $k=0$, $k=2$, and recoil-weighted $k=1$, $k=3$ multipoles of the unlike-nucleon spin-orbit interaction. Defining
\begin{align}
U_k^{(ls)}(R) &=
\int_0^\infty dr_t\,
u_{l_t}^2(r_t)\,
v_k^{(ls)}(R,\alpha r_t),
\\
U_{1/r}^{(ls)}(R) &=
\int_0^\infty dr_t\,
\frac{u_{l_t}^2(r_t)}{r_t}\,
v_1^{(ls)}(R,\alpha r_t),
\\
U_{3/r}^{(ls)}(R) &=
\int_0^\infty dr_t\,
\frac{u_{l_t}^2(r_t)}{r_t}\,
v_3^{(ls)}(R,\alpha r_t),
\end{align}
the radial combinations used to assemble the form factors quoted here are
\begin{align}
W_0^{(ls)}(R) &= \alpha U_0^{(ls)}(R) - R\,U_{1/r}^{(ls)}(R),
\label{eq:W0ls}
\\
W_2^{(ls)}(R) &= \alpha U_2^{(ls)}(R) - R\,U_{1/r}^{(ls)}(R),
\label{eq:W2ls}
\\
W_3^{(ls)}(R) &= \alpha U_2^{(ls)}(R) - R\,U_{3/r}^{(ls)}(R).
\label{eq:W3ls}
\end{align}
These equations define the spin-orbit radial combinations used in the
tensor/spin-orbit-included form-factor set. The corresponding
spin-orbit contribution to the form factor is written as
\begin{equation}
F_{ik}^{(ls)}(R)
=
\sum_{m\in\{0,2,3\}} C_{ikm}^{(ls)}(I,l_t)\,W_m^{(ls)}(R),
\label{eq:Fik_ls_appendix}
\end{equation}
where the channel-dependent coefficients $C_{ikm}^{(ls)}$ are obtained from the formula in Refs.~\cite{McAbee1990,McAbee1986}. In the present stretched-channel application those coefficients are made explicit directly in the assembled $F_{10}$, $F_{12}$, and $F_{32}$ expressions below.

\subsection{\texorpdfstring{Assembled $^{93}$Nb form factors}{Assembled 93Nb form factors}}

For the stretched $g_{9/2}$ channel of $^{93}$Nb, the coefficients used here are 
$a=1/4,~b=1$,$~c=-2/11,$ and 
the coefficients in $F_{32}$ are reduced to 
$c_{32}^{(ss)}=-3/11,~c_{32}^{(22)}=6/77,~c_{32}^{(24)}=162/1001$, and $c_{32}^{(ls)}=9/11$. 
The assembled form-factor sets used in the present analysis are

\begin{widetext}
\begin{align}
{F_{10}^{({ss})}}(R)
&=
a\, U_0^{(ss)}(R)~,~ {F_{10}}(R)={F_{10}^{({ss+t+ls})}}(R)
=
a\, U_0^{(ss)}(R)
+ c\, T_{02}(R)
- b\,W_0^{(ls)}(R),
\label{eq:Fik_assembled_f10_appendix}
\\
{F_{12}^{({ss})}}(R)
&=
c\, U_2^{(ss)}(R)~,~
{F_{12}}(R)={F_{12}^{({ss+t+ls})}}(R)
=
c\, U_2^{(ss)}(R)
- c\, T_{22}(R)
+ 2a\, T_{20}(R)
+ b\,W_2^{(ls)}(R),
\label{eq:Fik_assembled_f12_appendix}
\\
{F_{32}^{({ss})}}(R)
&=
\frac{3}{2}\,{F_{12}^{({ss})}}(R)~,~
{F_{32}}(R)={F_{32}^{({ss+t+ls})}}(R)
=
c_{32}^{(ss)}\, U_2^{(ss)}(R)
+ c_{32}^{(22)}\, T_{22}(R)
+ c_{32}^{(24)}\, T_{24}(R)
+ c_{32}^{(ls)}\,W_3^{(ls)}(R),
\label{eq:Fik_assembled_appendix}
\end{align}
\end{widetext}

Equations~(\ref{eq:Fik_assembled_f10_appendix})--(\ref{eq:Fik_assembled_appendix}) give the $F^{({ss})}$ and
$F^{({ss+t+ls})}$ sets shown in Fig.~\ref{fig:formfactors}.

\begin{acknowledgments}
K.~Heo and M.-K.C. acknowledge support from the National Research Foundation of Korea (NRF) under Grant No. RS-2024-00460031. 
M.-K.C. was also supported by the NRF Basic Science Research Program under Grant Nos. RS-2021-NR060129 and RS-2025-16071941. 
This work was also supported in part by
JSPS KAKENHI Grant Number JP23K03414. 
\end{acknowledgments}

\bibliographystyle{apsrev4-2}
\bibliography{references}

\end{document}